# Efficient multi-class fetal brain segmentation in high resolution MRI reconstructions with noisy labels

Kelly Payette[1], Raimund Kottke[2], Andras Jakab[1]

[1] Center for MR-Research, University Children's Hospital Zurich, Zurich, Switzerland,
kelly.payette@kispi.uzh.ch
[2] Diagnostic Imaging and Intervention, University Children's Hospital Zurich, Zurich Switzerland

**Abstract.** Segmentation of the developing fetal brain is an important step in quantitative analyses. However, manual segmentation is a very time-consuming task which is prone to error and must be completed by highly specialized individuals. Super-resolution reconstruction of fetal MRI has become standard for processing such data as it improves image quality and resolution. However, different pipelines result in slightly different outputs, further complicating the generalization of segmentation methods aiming to segment super-resolution data. Therefore, we propose using transfer learning with noisy multi-class labels to automatically segment high resolution fetal brain MRIs using a single set of segmentations created with one reconstruction method and tested for generalizability across other reconstruction methods. Our results show that the network can automatically segment fetal brain reconstructions into 7 different tissue types, regardless of reconstruction method used. Transfer learning offers some advantages when compared to training without pre-initialized weights, but the network trained on clean labels had more accurate segmentations overall. No additional manual segmentations were required. Therefore, the proposed network has the potential to eliminate the need for manual segmentations needed in quantitative analyses of the fetal brain independent of reconstruction method used, offering an unbiased way to quantify normal and pathological neurodevelopment.

## 1 Introduction

Fetal MRI is a useful modality for prenatal diagnostics and has a proven clinical value for the assessment of intracranial structures. Quantitative analysis potentially provides added diagnostic value and helps the understanding of normal and pathological fetal brain development. However, the quantitative assessment of fetal brain volumes requires accurate segmentation of fetal brain tissues, which can be a challenging task. Artifacts resulting from fetal and maternal movement are present, leading to difficulty in differentiating tissue types. Recently, advances have been made in the processing and super-resolution (SR) reconstruction of motion-corrupted low resolution fetal brain scans into high resolution volumes [1–7]. The enhanced resolution and improved image quality of the SR data in comparison to the native low-resolution scans has in turn resulted in greatly improved volumetric fetal brain data, the segmentation of which has not been assessed in detail.

Segmentation of the fetal brain from maternal tissue in the original scans has been explored [3, 8–11]. Methods to segment the original low resolution MR scans into different brain tissues have also been evaluated [12], as well as for a single tissue type within a high resolution volume [13, 14]. MRI atlases of the fetal brain have been generated with the intent of being used for atlas based-segmentation methods [9, 15, 16]. However, atlas-based methods are not easily expandable to pathological fetal brains, as currently no publicly available pathological fetal brain atlases exist. To overcome this, we propose using a multi-class U-Net for the segmentation of different types of fetal brain tissues in high resolution SR volumes.

The network should be able to work with SR reconstructions, regardless of the method used to create the fetal brain volume without requiring new manual segmentations. This can be challenging, as there are differences in shape, structure boundaries, textures, and intensities between volumes created by different reconstruction methods. Therefore, a network trained with one reconstruction method is not necessarily generalizable to other SR reconstruction methods, even when the same input data is used. To overcome this, the original labels can be rigidly registered to the alternate SR volume, creating 'noisy' labels, where noisy labels refer to incorrect labelling of the fetal brain volume as opposed to noise or artifact within the image itself. A network can then be trained using these noisy labels, thereby eliminating the need for further time consuming, manual brain segmentations.

Noisy labels have been shown to be a challenge for neural networks, where more noise results in performance degradation [17, 18]. Considerable research has been devoted to developing effective methods for handling noise, such as transfer learning, alternate loss functions, data re-weighting, changes to network architecture, and label cleaning, among others [19]. As the proposed method falls under the category of 'same task, different domain', transfer learning will be explored, as well as an alternate loss function (mean absolute error, MAE) that has been shown to be robust in the presence of noisy labels [20, 21]. Through transfer learning, the weights from the first network with 'clean' labels will be used for the initialization of a network with noisy labels, providing an automatic, objective segmentation of the fetal brains that is independent of SR method used.

Our proposed method aims to overcome these limitations (anatomical variability, the challenge of generalizability across SR methods, and the lack of noise-free, unambiguous anatomical annotations) and allow for multi-class fetal brain tissue segmentation across multiple SR reconstruction methods. Improvements to SR reconstruction methods can then be easily utilized for the quantitative analysis of the development of fetal brains with no new time-consuming manual segmentations required.

## 2 Methods

### 2.1 Image Acquisition

Multiple low-resolution orthogonal MR sequences of the brain were acquired on 1.5T and 3T clinical GE whole-body scanners at the University Children's Hospital Zurich using a T2-weighted single-shot fast spin echo sequence (ssFSE), with an in-plane resampled resolution of 0.5x0.5mm and a slice thickness of 3-5mm for 15 subjects. Each subject underwent a fetal MRI for a clinical indication and were determined to have unaffected neurodevelopment. The average gestational age in weeks (GA) of the subjects at the time of scanning was 28.7 ± 3.5 weeks (range: 22.6 - 33.4 GA).

### 2.2 Super-resolution Reconstruction

For each subject's set of images, SR reconstruction was performed using three different methods: mialSRTK [4], Simple IRTK [1], and NiftyMIC [3] using the following steps:
<u>Preprocessing:</u> The acquired images were bias corrected and de-noised prior to reconstruction using the tools included within each pipeline where applicable.
<u>Masking:</u> Each reconstruction method had different masking requirements. For the mialSRTK method, we reoriented and masked the fetal brains in each low-resolution image using a semi-automated atlas-based custom MeVisLab module [22]. For Simple IRTK and NiftyMIC, re-orientation of the input images was not required. For Simple IRTK, a brain mask was needed for the reference low-resolution image only, and this was generated using the network from [11], re-trained on the masks created with the aforementioned MeVisLab module. For NiftyMIC, the masking method available within the software was used.
<u>SR Reconstruction:</u> After pre-processing and masking, each SR reconstruction method was performed for each subject's low-resolution scans (mialSRTK, Simple IRTK, and NiftyMIC), resulting in three different fetal brain SR reconstructions with a resolution of 0.5x0.5x0.5mm, created with the same set of low-resolution input scans. Each SR volume was rigidly registered to the atlas space using FSL's flirt [23]. See Fig. 1 for examples of each SR reconstruction.

### 2.3 Image Segmentation

A 2D U-Net was chosen as the basis for image segmentation [24] with an adam optimizer, a learning rate of 10E-5, L2 regularization, ReLu activation, and batch normalization after each convolutional layer, and a dropout layer after each block of convolutional layers. The network was programmed in Keras and trained on an Nvidia Quadro P6000. The network was trained for 100 epochs with early stopping. The training data was histogram-matched and normalized prior to training. The 2D U-Net was trained in the axial orientation. The generalized Dice coefficient and the mean absolute error (MAE) were used as a loss functions [25]. The initial network was trained on a set of n images $\{r_i, l_i\}$, where $r_i$ is a reconstructed image using the mialSRTK method, and $l_i$ is the corresponding manually annotated label map (The mialSRTK method was chosen due to availability of manual annotations). In order to create $l_i$, the SR volume using the mialSRTK method was segmented into 7 tissue types (white matter (WM), grey matter (GM), external cerebrospinal fluid (eCSF), ventricles, cerebellum, deep GM, brain stem). Two volumes (GA: 26.6, 31.2) were retained for validation. The remaining 13 volumes were used for training and testing. Data augmentation (flipping, 360° rotation, adding Gaussian noise) was also utilized. In addition, we registered the manual label maps to the Simple IRTK and NiftyMIC SR volumes using ants and an age-matched label map in order to compare a simple atlas-based method to the U-Net [26].

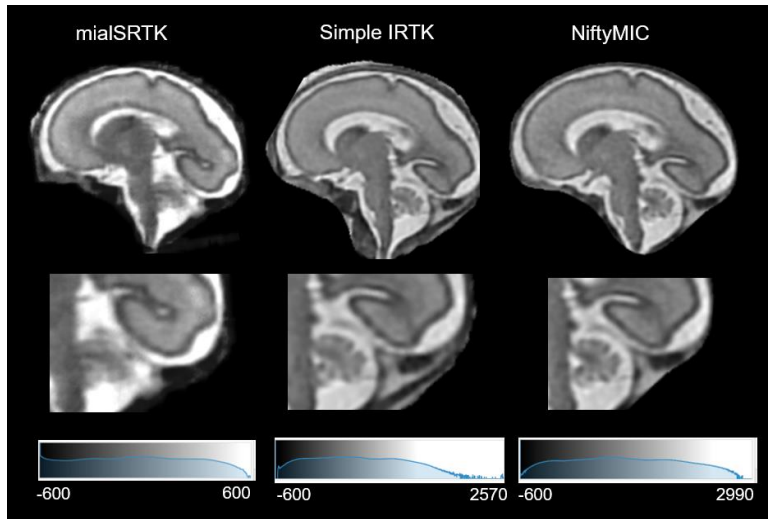

**Fig. 1.** SR reconstructions with each method. Top row: Complete SR reconstruction; middle row: enlarged area of the SR method; bottom row: intensity histogram of each reconstruction (27.7 GA). Each method has variations in shape, structure boundaries, texture, image contrast, and they retain different amounts of non-brain tissue for the same subject.

For each alternate reconstruction method, a set of n images $\{z_i', l_i\}$ was generated, where $z_i$ denotes the $i^{th}$ reconstruction created with the alternate method, and $l_i$ denotes the original labels. The noisy labels were created by rigidly registering the new reconstruction ($z_i'$) to the existing label map using flirt [23]. Errors in the registration, plus the difference between the reconstruction methods cause the original labels to only be an approximate match to the new reconstruction (the so-called 'noisy' labels), as shown in Fig. 2.

The weights generated in the initial mialSRTK U-Net (Network 1) were used as weight initializations for training segmentation networks for the other reconstruction methods. See Table 1 for a detailed overview of the networks trained.

In addition, volumes created with NiftyMIC and Simple IRTK were segmented with Network 1 as comparison. The volumes retained for validation in Simple IRTK and NiftyMIC were manually annotated for validation purposes. The networks were evaluated by comparing the individual labels of the newly segmented label and the original annotation using the Dice coefficient (DC) and the 95% percentile of the Hausdorff distance (HD) [27].

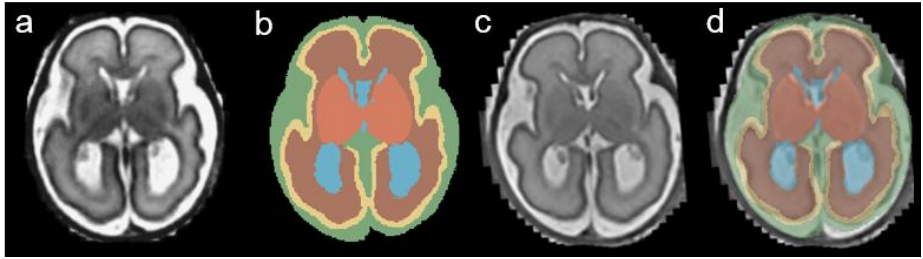

**Fig. 2.** a) mialSRTK reconstruction; b) manually annotated label; c) Simple IRTK reconstruction registered to the original mialSRTK image and corresponding label; d) the label displayed in b) overlaid on the SR volume shown in c), showing that the same label overlaid on Simple IRTK volume is noisy, leading to mislabeling in the anterior GM and ventricles, and posterior external CSF space in this slice.

**Table 1.** Overview of the networks. Note: networks 2-6 were created for each alternate SR method (Simple IRTK, NiftyMIC)

| Network Number | Images | Labels | Weight Initialization | Loss Function |
|---|---|---|---|---|
| 1 | $r_i$: SR volume created with mialSRTK | $l_i$ | glorot uniform | Generalized Dice |
| 2 | $z_i$': SR volume created with alternate SR method, registered to labels $l_i$ | $l_i$ | glorot uniform | Generalized Dice |
| 3 | $z_i$': SR volume created with alternate SR method, registered to labels $l_i$ | $l_i$ | Transfer Learning (from Network 1) | Generalized Dice |
| 4 | $z_i$': SR volume created with alternate SR method, registered to labels $l_i$ | $l_i$ | glorot uniform | MAE |
| 5 | $z_i$': SR volume created with alternate SR method, registered to labels $l_i$ | $l_i$ | Transfer Learning (from Network 1) | MAE |
| 6 | $r_i$ and $z_i$', registered to labels $l_i$ | $l_i$ | glorot uniform | Generalized Dice |

## 3 Results

The network with the original labels and SR method 1 (mialSRTK) performs with an average Dice coefficient of 0.86 for all tissue types, with values ranging from 0.672 (GM) to 0.928 (cerebellum). When the SR methods 2 (Simple IRTK) and 3 (NiftyMIC) are run through the same network (Network 1), all labels perform on average 0.04-0.05 DC points lower, while the HD results seem to vary label to label (see Fig. 4, 5). Interestingly, the ventricles and cerebellum in the Simple IRTK SR reconstruction are segmented more accurately than in the mialSRTK volume, potentially due to stronger intensity differences between tissue and CSF.

The transfer learning is a clear improvement on training the noisy labels from a standard weight initialization (from an average DC of 0.62 to 0.79 with the generalized dice loss function, as well as improving the average HD from 34.4 to 25.2), but it fails to outperform the original network (based on the average DC: 0.81). Using the MAE loss with the transfer learning is not as accurate as with the generalized Dice coefficient loss (average DC: 0.70, average HD: 26.6) as the network is unable to classify the brainstem in one of the SR methods. It is also unable to detect all required classes in both alternate SR methods when trained without transfer learning. The is potentially due to the class imbalance (the network fails to find the smaller classes by number of voxels such as the brainstem and cerebellum but can detect the larger classes such as GM and WM). Within each SR volume method, some tissue classes are segmented more accurately than those within the reference SR volume trained in Network 1, even if the average across all tissues for each network is lower. Combining all volumes together in one network training set (Network 6) resulted in high dice scores in the cerebellum, deep GM, and the brainstem, but the overall average of the Dice scores across all labels was lower when compared to other networks. The ants registration segmentation did not perform as well when looking at the DC (average DC: 0.78), but it drastically out-performed all of the networks when looking at the HD (average HD: 14.7). Example segmentations of each SR method can be found in Fig. 3.

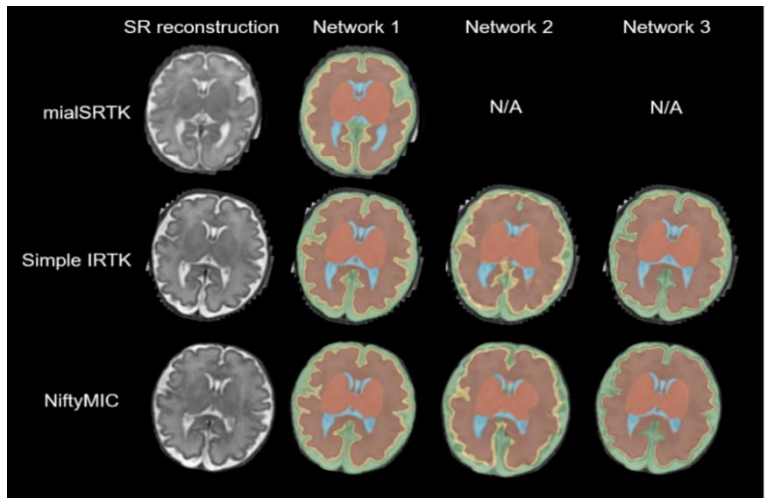

**Fig. 3.** Automatic segmentations created by Networks 1, 2, and 3 for each SR method. Network 2 (noisy labels without transfer learning) has difficulty delineating the GM, and the midline is shifted. These segmentation errors are resolved in Network 3 (with transfer learning), however the cortex is thinner.

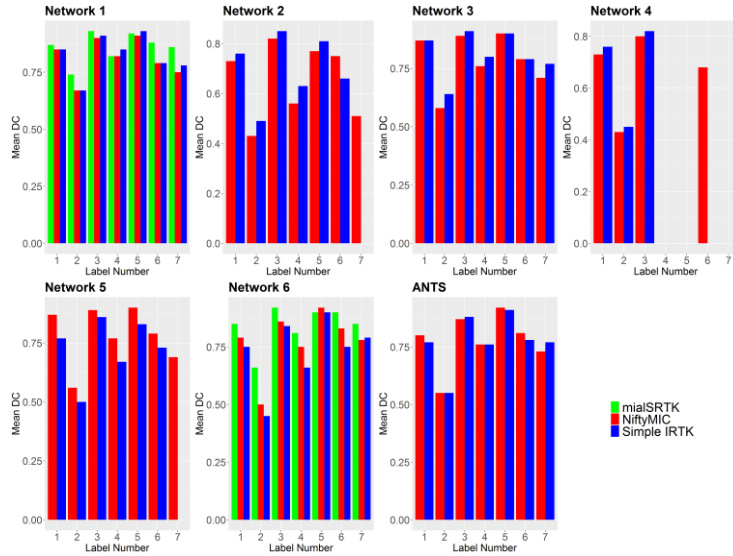

**Fig. 4** Average DC for each label in each network. The mialSRTK volumes have the highest scoring DC. The networks that use transfer learning (3 and 5) perform as well as Network 1 for SR methods NiftyMIC and Simple IRTK, but do not reach the same value as mialSRTK in Network 1. There was no overlap for labels 4, 5, and 7 in the NiftyMIC method, and none for labels 4-7 in Simple IRTK in Network 4.

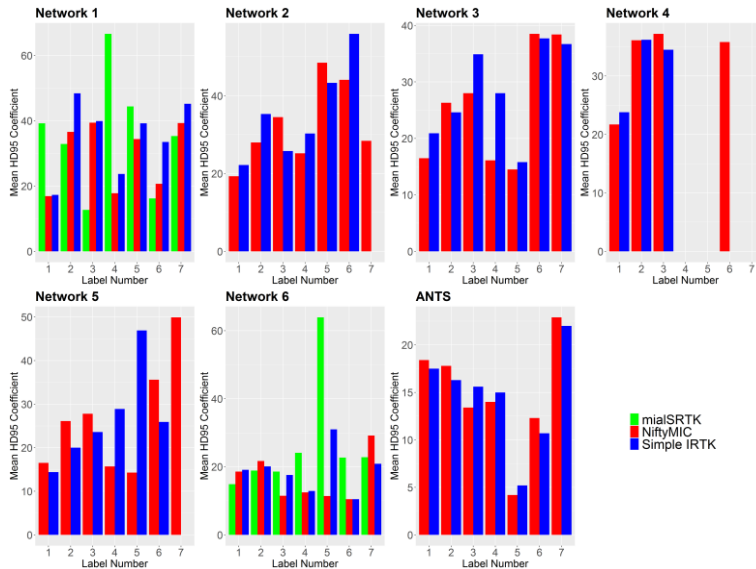

**Fig. 5** Average HD (95 percentile) for all classes for each network.

## 4      Discussion and Conclusion

In this research we showed that the automatic segmentation of fetal brain volumes can be generalized across different SR reconstruction methods. Segmentation accuracy in the presence of noisy labels is very challenging, and can be helped with transfer learning, although not to the level of a network trained on clean labels. High resolution fetal brain volumes created from three distinct methods were able to be segmented into 7 different tissue types using a U-Net trained with transfer learning and noisy labels. Transfer learning can increase the quality of the segmentation across SR methods, but cannot outperform training a model on clean labels. The choice of loss function is important, especially when training a network without pre-initializing the weights. The MAE loss function does not perform as well as the generalized dice loss function in the presence of noisy labels. A potential improvement to the model would be to expand it to a 3D model, which may potentially improve the segmentation accuracy, but would require either increased data augmentation or a larger training dataset. We expect further increase of segmentation performance after including additional cases, potentially representing broader gestational age range and larger anatomical variability including pathological fetal brains, thus increasing the generalizability of our approach. The atlas-based segmentation method performed incredibly well when looking at the HD values. This is potentially because in atlas-based segmentations, existing shape data exists, which improves the shape of the segmentation, even if the overlap is not as correct as in the U-Net. However, the SR reconstructions used here are considered to be neurodevelopmentally normal brains, so the effectiveness of this method is unknown for pathological brain structures.

This method will limit the amount of manual segmentation needed for future quantitative analyses of fetal brain volumes. It could also potentially be used as an automated segmentation training strategy for when new SR algorithms are developed in the future. In addition, this could potentially be used to further investigate the differences between the various SR methods in order to understand the quantitative differences between the methods, and how the method chosen could impact analyses. In the future, this method can potentially be used to expand the network's applicability for use with data from other MR scanners, across study centers, or with new SR algorithms.

**Acknowledgements**

Financial support was provided by the OPO Foundation, Anna Müller Grocholski Foundation, the Foundation for Research in Science and the Humanities at the University of Zurich, EMDO Foundation, Hasler Foundation, the Forschungszentrum für das Kind Grant (FZK) and the PhD Grant from the Neuroscience Center Zurich.